\documentclass[prd,showpacs,showkeys,nofootinbib,floatfix,
               fleqn,preprint,12pt,tightenlines]{revtex4-1}

\usepackage{amsmath,amssymb,revsymb,graphicx,dcolumn}
\usepackage{array,hyperref}
\usepackage{dcolumn}  

\newcommand{\beq}{\begin{equation}}
\newcommand{\eeq}{\end{equation}}
\newcommand{\beqa}{\begin{eqnarray}}
\newcommand{\eeqa}{\end{eqnarray}}
\newcommand{\bsubeqs}{\begin{subequations}}
\newcommand{\esubeqs}{\end{subequations}}

\newcommand{\half}{{\textstyle \frac{1}{2}}}

\begin{document}

\begin{widetext}
\noindent Acta Phys. Polon. B \textbf{54}, 5-A3 (2023) \hfill  arXiv:2301.00724
%
%
\newline\vspace*{0mm}
\end{widetext}

\title{Defect wormhole: A traversable wormhole without exotic matter\vspace*{4mm}}

 \author{F.R. Klinkhamer}
 \email{frans.klinkhamer@kit.edu}
 \affiliation{Institute for Theoretical Physics,
 Karlsruhe Institute of Technology (KIT),\\
 76128 Karlsruhe, Germany\\}

\vspace*{10mm}

\begin{abstract}
\vspace*{2.5mm}\noindent
We present a traversable-wormhole solution of the
gravitational field equation of General Relativity
without need of exotic matter
(exotic matter can, for example, have negative energy density
and vanishing isotropic pressure).  
Instead of exotic matter, the solution relies on a 3-dimensional
``spacetime defect'' characterized by a locally vanishing metric
determinant.
\vspace*{10mm}
\end{abstract}


\maketitle

\section{Introduction}
\label{sec:Introduction}

Traversable wormholes~\cite{MorrisThorne1988}
appear to require ``exotic'' matter,
for example matter violating the Null Energy Condition (NEC).
See, e.g., Ref.~\cite{Visser1996} for further discussion and
references.

In this paper, we look for a way around the necessity
of having exotic matter, while making no essential changes
in the established theories (General Relativity 
and the Standard Model of elementary particle physics).

Throughout, we use natural units with $c=1$  and $\hbar=1$.

\section{Basic idea}
\label{sec:Basic-idea}

The regularized-big-bang spacetime~\cite{Klinkhamer2019a,Klinkhamer2019b}
is a solution of the gravitational field equation of
General Relativity with normal matter and a degenerate metric.
This spacetime corresponds to
a traversable cosmic bounce~\cite{KlinkhamerWang2019a,KlinkhamerWang2019b}
(a brief review appears in  Ref.~\cite{Klinkhamer2021-review}).
For comments on the standard version of General Relativity
and the extended version used here,
see the last two paragraphs in Sec.~I of Ref.~\cite{Klinkhamer2019a}.

As noted briefly in Sec.~II of Ref.~\cite{Klinkhamer2019b}
and more extensively in Sec.~II of Ref.~\cite{KlinkhamerWang2019b},
the degeneracy of the regularized-big-bang
metric gives an effective matter component
which is ``exotic,'' specifically NEC violating.
(The NEC~\cite{Visser1996}
corresponds to the following requirement 
on the energy-momentum tensor $T^{\,\mu\,\nu}$ for an
arbitrary null-vector $k_{\mu}\,$:
$T^{\,\mu\,\nu}\,k_{\mu}\,k_{\nu} \geq 0$.)
The heuristics, then, is that the exotic effects of the metric degeneracy
turn the singular (concave) big-bang behavior
[\,$a(t) \sim \sqrt{t} \to 0$ for $t \downarrow 0$]
into a smooth (convex) bounce
behavior [$a(T) \sim a_{B}+T^{2}$   
for $T \in (- \Delta T,\,+\Delta T)$, with $a_{B} > 0$ and
$\Delta T > 0$\,].

We now try to do something similar for the traversable wormhole,
using General Relativity with normal matter but allowing
for a degenerate metric.

\section{Simple example}
\label{sec:Simple-example}

\subsection{Nondegenerate metric -- special case}
\label{subsec:Nondegenerate-metric-special-case}

We can test the basic idea of Sec.~\ref{sec:Basic-idea}
if we start from the simple example discussed
by Morris and Thorne (MT) in  Box~2 of Ref.~\cite{MorrisThorne1988}.
There, the special case of a more general metric
is given by (recall $c=1$)
\beqa
\label{eq:EBMT-special}
ds^{2}\,\Big|^\text{(EBMT-worm-spec)}
&\equiv&
g_{\mu\nu}(x)\, dx^\mu\,dx^\nu \,\Big|^\text{(EBMT-worm-spec)}
\nonumber\\[1mm]
&=&
- dt^{2} + dl^{2}
+ \left(b_{0}^{2} + l^{2}\right)\,
  \Big[ d\theta^{2} + \sin^{2}\theta\, d\phi^{2} \Big]\,,
\eeqa
with a nonzero real constant $b_{0}$ (taken to be positive, for definiteness).
The coordinates $t$ and $l$ in \eqref{eq:EBMT-special} 
range over $(-\infty,\,\infty)$, and the coordinates  $\theta \in [0,\,\pi]$ 
and $\phi \in [0,\,2\pi)$ are the standard spherical polar coordinates 
(see the paragraph below for a technical remark).
Earlier discussions of this type of metric
have appeared in the independent papers of Ellis~\cite{Ellis1973}
and Bronnikov~\cite{Bronnikov1973},
and, for this reason, we have added  ``EB''
to the suffix in \eqref{eq:EBMT-special}.

The announced technical remark,
which can be skipped in
a first reading, is about the coordinates of the 2-sphere.
Instead of the single set $\{\theta,\,\phi\}$,
we should really use two (or more) appropriate coordinate patches
for the 2-sphere~\cite{EguchiGilkeyHanson1980,Nakahara2003}. 
A well-known example has
the coordinates $\{X,\,Y\}$ obtained by stereographic projection
from the North Pole ($\theta=0$) on the equatorial plane $\mathbb{R}^{2}$
and the coordinates $\{U,\,V\}$ obtained by stereographic projection
from the South Pole ($\theta=\pi$); see Exercise 5.1 in Sec.~5.1
of Ref.~\cite{Nakahara2003}.
For the coordinates of the first patch, the last term in
the squared line element from \eqref{eq:EBMT-special}
is replaced by
$4\left(b_{0}^{2} + l^{2}\right) (1+X^{2}+Y^{2})^{-2}\:
\big[ dX^{2} +  dY^{2} \big]$ and,
for the coordinates of the second patch, there is the term
$4\left(b_{0}^{2} + l^{2}\right) (1+U^{2}+V^{2})^{-2}\:
\big[ dU^{2} +  dV^{2} \big]$.
In the first coordinate patch,
the resulting metric components $g_{XX}$ and $g_{YY}$
vanish nowhere, and similarly for $g_{UU}$ and $g_{VV}$ in the second patch
[by contrast,
the metric component $g_{\phi\phi}$ from \eqref{eq:EBMT-special}
vanishes at two points, $\theta=0$ and $\pi$, with
$\sin^{2}\theta=0$].

Let us continue the discussion of the
metric \eqref{eq:EBMT-special} as it stands. Then,
according to items (d) and (e) of Box~2 in Ref.~\cite{MorrisThorne1988},
the wormhole from \eqref{eq:EBMT-special}
is traversable; see also Fig.~6 in Ref.~\cite{Ellis1973}.

The crucial question, however, is the dynamics:
can the wormhole metric \eqref{eq:EBMT-special}
be a solution of the Einstein equation?
Morris and Thorne's used a type of engineering approach:
fix the desired specifications and see what it takes.
The Einstein equation,
$G_{\mu\nu} \equiv R_{\mu\nu} - \half\, g_{\mu\nu}\,R = 8\pi G\:T_{\mu\nu}$,
then requires the following
components of the energy-momentum tensor~\cite{MorrisThorne1988}:
\bsubeqs\label{eq:T-components-EBMT-special}
\beqa
\label{eq:T-tt-component-EBMT-special}
T^{\,t}_{\;\;\,t}\,\Big|^\text{(EBMT-worm-spec)}
&=& \frac{1}{8\pi G}\;
\frac{b_{0}^{2}}{\left(b_{0}^{2} + l^{2}\right)^{2}}\,,
\\[2mm]
T^{\,l}_{\;\;\,l}\,\Big|^\text{(EBMT-worm-spec)}
&=& -\frac{1}{8\pi G}\;
\frac{b_{0}^{2}}{\left(b_{0}^{2} + l^{2}\right)^{2}}\,,
\\[2mm]
T^{\,\theta}_{\;\;\,\theta}\,\Big|^\text{(EBMT-worm-spec)}
&=& \frac{1}{8\pi G}\;
\frac{b_{0}^{2}}{\left(b_{0}^{2} + l^{2}\right)^{2}}\,,
\\[2mm]
T^{\,\phi }_{\;\;\,\phi}\,\Big|^\text{(EBMT-worm-spec)}
&=& \frac{1}{8\pi G}\;
\frac{b_{0}^{2}}{\left(b_{0}^{2} + l^{2}\right)^{2}}\,,
\eeqa
\esubeqs
with all other components vanishing.

The energy density is given by $\rho = T^{\,tt} =-T^{\,t}_{\;\;\,t}$ and
we have
$\rho< 0$ from \eqref{eq:T-tt-component-EBMT-special},
which definitely corresponds to unusual matter.
Moreover, we verify, for the radial null vector
$\widehat{k}^{\,\mu}=(1,\, 1,\,  0,\,0)$,  
the inequality
\beq
\label{eq:EBMT-special-NEC-radial}
T^{\,\mu}_{\;\;\,\nu}\,
\widehat{k}_{\mu}\,\widehat{k}^{\,\nu}\,\Big|^\text{(EBMT-worm-spec)}
= \frac{1}{8\pi G}\;
\frac{b_{0}^{2}}{\left(b_{0}^{2} + l^{2}\right)^{2}}\;\big[-1-1\big] < 0\,,
\eeq
which corresponds to NEC violation.
For tangential null vectors, we get the same expression
\eqref{eq:EBMT-special-NEC-radial},
but now with the factor $[-1+1]=0$.

\subsection{Degenerate metric -- special case}
\label{subsec:Degenerate-metric-special-case}

We, next, consider the following metric \emph{Ansatz}:
\beq
\label{eq:K-special}
ds^{2}\,\Big|^\text{(K-worm-spec)}
=
- dt^{2} + \frac{\xi^{2}}{\lambda^{2} + \xi^{2}}\;d\xi^{2}
+ \left(b_{0}^{2} + \xi^{2}\right)\,
  \Big[ d\theta^{2} + \sin^{2}\theta\, d\phi^{2} \Big]\,,
\eeq
with nonzero real constants $\lambda$ and $b_{0}$
(both taken to be positive, for definiteness)
and coordinates $t$ and $\xi$ ranging over $(-\infty,\,\infty)$.
The metric from \eqref{eq:K-special} gives
the following Ricci and Kretschmann curvature scalars:
\bsubeqs\label{eq:K-special-R-K}
\beqa
R\,\Big|^\text{(K-worm-spec)} &=&
-2\;\frac{b_{0}^{2}-\lambda^{2}}
         {\left(b_{0}^{2} + \xi^{2}\right)^{2}}\,,
\\[2mm]
K\,\Big|^\text{(K-worm-spec)} &=&
12\;\frac{\left(b_{0}^{2}-\lambda^{2}\right)^{2}}
         {\left(b_{0}^{2} + \xi^{2}\right)^{4}}\,,
\eeqa
\esubeqs
both of which are finite, perfectly smooth,
and vanishing for $\xi \to \pm\infty$.

The metric $g_{\mu\nu}(x)$ from \eqref{eq:K-special}
is degenerate with a vanishing determinant
$g(x)\equiv \det[g_{\mu\nu}(x)]$ at $\xi=0$.
[Note that the metric $g_{\mu\nu}(x)$ from \eqref{eq:EBMT-special} is
nondegenerate, as its determinant $g(x)$ vanishes nowhere,
provided two suitable coordinate patches
are used for the 2-sphere.]
In physical terms, this 3-dimensional hypersurface at $\xi=0$
corresponds to a ``spacetime defect''
\cite{Klinkhamer2014-prd,KlinkhamerSorba2014,%
Guenther2017,Klinkhamer2019-jpcs}
and the Einstein equation
is defined at $\xi=0$ by continuous extension from
its limit $\xi \to 0$  (for this last point, see, in particular,
Sec.~3.3.1 of Ref.~\cite{Guenther2017} 
and also the related discussion in  
Sec.~\ref{subsubsec:Vacuum-solution-first-order-equations}). 
The terminology ``spacetime defect'' is
by analogy with crystallographic defects in an atomic crystal
(these crystallographic defects are typically formed during
a rapid crystallization process).

We now have two further technical remarks,
which can be skipped in a first reading.
First, we might consider changing
the quasi-radial $\xi$ coordinate to
\beq
\label{eq:ltilde}
\widetilde{l} = \xi\;\sqrt{1+\lambda^{2}/\xi^{2}}
 \in (-\infty,\,-\lambda] \cup [\lambda,\,\infty)\,,
\eeq
which would give a metric similar to \eqref{eq:EBMT-special},
\beq
\label{eq:ltilde-metric}
ds^{2}=
- dt^{2} + d\widetilde{l}^{\;2}
+ \left(b_{0}^{2} + \widetilde{l}^{\;2} - \lambda^{2}\right)\,
  \Big[ d\theta^{2} + \sin^{2}\theta\, d\phi^{2} \Big]\,.
\eeq
But this coordinate transformation $\xi \to \widetilde{l}$
is discontinuous and, therefore, \emph{not} a diffeomorphism.
We also remark that the coordinate
$\widetilde{l}$ from \eqref{eq:ltilde}
is unsatisfactory for the correct description
of the \emph{whole} spacetime manifold,
as, for given values of $\{t,\,\theta,\,\phi\}$,
both $\widetilde{l}=-\lambda$ and $\widetilde{l}=\lambda$
correspond to a single point of the manifold (with the single
coordinate $\xi=0$).
The proper coordinates of the defect-wormhole
spacetime \eqref{eq:K-special}  
are $\{t,\, \xi,\, \theta,\,  \phi\}$ and
not $\{t,\, \widetilde{l},\, \theta,\,  \phi\}$,
or possible regularizations based on the latter coordinates.
For further discussion of some of
the physics and mathematics
issues of such spacetime defects, see
Sec.~III of Ref.~\cite{KlinkhamerSorba2014}
and Sec.~3 of Ref.~\cite{Guenther2017}.

Second, the embedding diagram
of the spacetime \eqref{eq:K-special}
for  $(t,\,\theta)=(\text{const},\,\pi/2)$
and $0 < \lambda^{2} < b_{0}^{2}$
is similar [with a 3-dimensional Euclidean embedding space]
to the embedding diagram
of the spacetime \eqref{eq:EBMT-special}
for the same values of $(t,\,\theta)$ and $b_{0}^{2}$,
as given by item (b) of
Box~2 in Ref.~\cite{MorrisThorne1988}.
The embedding diagram of the spacetime \eqref{eq:K-special}
for $\lambda^{2} > b_{0}^{2}$ is similar to the
embedding diagrams for $\lambda^{2} \in [0,\, b_{0}^{2})$,
except that, for $\lambda^{2} > b_{0}^{2}$, there is
a (2+1)-dimensional Minkowski embedding space.
These new embedding diagrams for
$\lambda^{2} > 0$ (and,  for the moment, $\lambda^{2} \ne b_{0}^{2}\,$)
are nonsmooth at $\xi=0$, which is a direct manifestation
of the presence of the spacetime defect.
In order to obtain smooth motion, we are led to nonstandard
identifications at the wormhole throat. This is especially
clear in the description  of the spacetime \eqref{eq:K-special}
at $\lambda^{2} = b_{0}^{2}$,
which has a flat metric \eqref{eq:ltilde-metric}
for $\lambda^{2} = b_{0}^{2}$ in terms of the auxiliary
quasi-radial variable $\widetilde{l}$.
The description then uses two copies of
the flat Euclidean space $E_3$ with the interior of
two balls with radius $\lambda$  removed
and their surfaces at $\widetilde{l}=\pm \lambda$
identified ``antipodally''
(see Sec.~\ref{subsec:Topology} for further details).

After these technical remarks, we return to
the metric \eqref{eq:K-special} and
observe that the Einstein equation
(defined at $\xi=0$ by the limit; see above) requires
\bsubeqs\label{eq:T-UPDOWNcomponents-degenmetric-special}
\beqa
\label{eq:T-UPtDOWNt-component-degenmetric-special}
T^{\,t}_{\;\;\,t}\,\Big|^\text{(K-worm-spec)}
&=& \frac{1}{8\pi G}\;
\frac{b_{0}^{2}-\lambda^{2}}{\left(b_{0}^{2}+\xi^{2}\right)^{2}}\,,
\\[2mm]
T^{\,\xi}_{\;\;\,\xi}\,\Big|^\text{(K-worm-spec)}
&=& -\frac{1}{8\pi G}\;
\frac{b_{0}^{2}-\lambda^{2}}{\left(b_{0}^{2}+\xi^{2}\right)^{2}}\,,
\\[2mm]
T^{\,\theta}_{\;\;\,\theta}\,\Big|^\text{(K-worm-spec)}
&=& \frac{1}{8\pi G}\;
\frac{b_{0}^{2}-\lambda^{2}}{\left(b_{0}^{2}+\xi^{2}\right)^{2}}\,,
\\[2mm]
T^{\,\phi }_{\;\;\,\phi}\,\Big|^\text{(K-worm-spec)}
&=& \frac{1}{8\pi G}\;
\frac{b_{0}^{2}-\lambda^{2}}{\left(b_{0}^{2}+\xi^{2}\right)^{2}}\,.
\eeqa
\esubeqs
Compared to the previous result \eqref{eq:T-components-EBMT-special}, 
we see that the previous factors $b_{0}^{2}$ in the numerators
have been replaced by new factors $(b_{0}^{2}-\lambda^{2})$,
with corresponding changes in the denominators
[\,$b_{0}^{2} \to b_{0}^{2}-\lambda^{2}$ and
$l^{2} \to \widetilde{l}^{\;2} =  \lambda^{2} + \xi^{2}$,
so that $b_{0}^{2}+l^{2} \to b_{0}^{2} + \xi^{2}$\,].
Starting from $\lambda^{2}=0^{+}$,
these new numerator factors $(b_{0}^{2}-\lambda^{2})$
then change sign as $\lambda^{2}$ increases above $b_{0}^{2}$  and
we no longer require exotic matter.

Indeed, we have from \eqref{eq:T-UPtDOWNt-component-degenmetric-special}
that $\rho =-T^{\,t}_{\;\;\,t} > 0$ for $\lambda^{2} > b_{0}^{2}\,$.
Moreover, we readily obtain,
for any null vector $k^{\,\mu}$
and parameters $\lambda^{2} \geq b_{0}^{2}\,$,
the inequality
\beq
\label{eq:EBMT-special-NEC}
T^{\,\mu}_{\;\;\,\nu}\,k_{\mu}\,k^{\,\nu}\,
\Big|^\text{(K-worm-spec)}_{\lambda^{2} \geq b_{0}^{2}}
\geq 0\,,
\eeq
which verifies the NEC [this result follows equally from the
expressions in Sec.~\ref{subsec:Nondegenerate-metric-special-case},
if we again replace the numerator factors
$b_{0}^{2}$ there by $(b_{0}^{2}-\lambda^{2})$ and make
corresponding changes in the denominators].

There is, of course, also the special case $\lambda^{2}=b_{0}^{2}$,
for which the energy-momentum tensor vanishes altogether,
\beq
\label{eq:T-UPmuUPnu-component-degenmetric-special}
T^{\,\mu}_{\;\;\,\nu}\,\Big|^\text{(K-worm-spec)}_{\lambda^{2}=b_{0}^{2}}
= 0\,,
\eeq
and so do the curvature scalars \eqref{eq:K-special-R-K}.
In that case, we have a defect wormhole in the vacuum, which will  
be discussed further in Sec.~\ref{subsec:Vacuum-solution},
where also the radial geodesics will be presented.

\section{Degenerate wormhole metric}
\label{sec:Degenerate wormhole metric}

\subsection{General Ansatz}
\label{subsec:General Ansatz}

The special degenerate metric \eqref{eq:K-special}
can be generalized as follows:
\beqa
\label{eq:K}
ds^{2}\,\Big|^\text{(K-worm-gen)}
&\equiv&
g_{\mu\nu}(x)\, dx^\mu\,dx^\nu \,\Big|^\text{(K-worm-gen)}
\nonumber\\[1mm]
&=&
- e^{2\,\widetilde{\phi}(\xi)}\;dt^{2}
+ \frac{\xi^{2}}{\lambda^{2} + \xi^{2}}\;d\xi^{2}
+ \widetilde{r}^{\;2}(\xi)\,
  \Big[ d\theta^{2} + \sin^{2}\theta\, d\phi^{2}  \Big]\,,
\eeqa
with a positive length scale $\lambda$ and
real functions $\widetilde{\phi}(\xi)$ and $\widetilde{r}(\xi)$.
Again, the coordinates $t$ and $\xi$
range over $(-\infty,\,\infty)$, while
$\theta \in [0,\,\pi]$ and $\phi \in [0,\,2\pi)$
are the standard spherical polar coordinates
[as mentioned in Sec.~\ref{subsec:Nondegenerate-metric-special-case},
we should really use two appropriate coordinate patches
for the 2-sphere].
If, moreover, we assume that
$\widetilde{\phi}(\xi)$ remains finite everywhere
and that $\widetilde{r}(\xi)$ is positive with
$\widetilde{r}(\xi) \sim |\xi |$ for $\xi \to \pm\infty$,
then the spacetime from
\eqref{eq:K} corresponds to a wormhole
(see also the discussion at the beginning of Sec.~11.2
in Ref.~\cite{Visser1996}).

If the global minimum of the function $\widetilde{r}(\xi)$
has the value $b_{0}>0$ at $\xi= 0$ and if  
the function $\widetilde{\phi}(\xi)$
is essentially constant near $\xi=0$,
then we expect interesting behavior
for $\lambda^{2}$  of the order of $b_{0}^{2}$ or larger.
In fact, using power series in $\xi^{2}$ for the \textit{Ansatz} functions
of the metric \eqref{eq:K} [specifically,
$\widetilde{\phi}(\xi)=c_{0}+c_{2}\,\xi^{2}+c_{4}\,\xi^{4}+ \ldots\,$
and
$\widetilde{r}^{\;2}(\xi)=b_{0}^{2} +d_{2}\,\xi^{2}+d_{4}\,\xi^{4}+ \ldots\,$],
we get energy-momentum components without singular
behavior at $\xi=0$. It is clear that further
work will be cumbersome but perhaps not impossible.
Some preliminary numerical results 
will be discussed at the end of Sec.~\ref{sec:Discussion}.

For later use, we already give the tetrad 
$e^{a}_{\phantom{z}\mu}$ corresponding to the general metric 
$g_{\mu\nu}=\eta_{ab}\,e^{a}_{\phantom{z}\mu}\,e^{b}_{\phantom{z}\nu}$
from \eqref{eq:K}:
\bsubeqs\label{eq:generalized-wormhole-tetrad}
\beqa
e^{0}_{\phantom{z}\mu}(x)\,\Big|^\text{(K-worm-gen)}
&=& e^{\widetilde{\phi}(\xi)}\;\delta^{0}_{\phantom{z}\mu}\,,
\\[2mm]
\label{eq:generalized-wormhole-tetrad-a-is-1}
e^{1}_{\phantom{z}\mu}(x)\,\Big|^\text{(K-worm-gen)}
&=&  \frac{\xi}{\sqrt{\lambda^{2} + \xi^{2}}}\;\delta^{1}_{\phantom{z}\mu}\,,
\\[2mm]
e^{2}_{\phantom{z}\mu}(x)\,\Big|^\text{(K-worm-gen)}
&=&  \sqrt{\widetilde{r}^{\;2}(\xi)}\;\delta^{2}_{\phantom{z}\mu}\,,
\\[2mm]
e^{3}_{\phantom{z}\mu}(x)\,\Big|^\text{(K-worm-gen)}
&=& \sqrt{\widetilde{r}^{\;2}(\xi)}\;\sin\theta \;\delta^{3}_{\phantom{z}\mu}\,,
\eeqa
\esubeqs
where the argument $x$ of the tetrad stands for the coordinates
$\big(t,\, \xi,\, \theta,\,  \phi\big)$. The particular
choice for \eqref{eq:generalized-wormhole-tetrad-a-is-1}
will be commented on in Sec.~\ref{subsubsec:Vacuum-solution-geodesics}.

\subsection{Topology and orientability}
\label{subsec:Topology}

For a brief discussion of the topology of
the spacetime with metric \eqref{eq:K}, we can set
$\widetilde{\phi}(\xi)=0$ and
$\widetilde{r}^{\;2}(\xi)=b_{0}^{2}+\xi^{2}$,
so that we are back to the special metric \eqref{eq:K-special}.
Then, from the auxiliary 
coordinates $\big\{\widetilde{l},\,\theta ,\, \phi\big\}$
in the metric \eqref{eq:ltilde-metric} for general $\lambda>0$
and $b_{0}>0$, we get the following two sets of
Cartesian coordinates
(one for the ``upper'' universe with $\widetilde{l}>\lambda$
and the other for the ``lower'' universe with $\widetilde{l}<-\lambda$):
\bsubeqs\label{eq:Cartesian-coordinates}
\beqa
\label{eq:Cartesian-coordinates-plus}
\left\{
\begin{array}{c}
  Z_{+} \\
  Y_{+} \\
  X_{+}
\end{array}
 \right\}
&=& \widetilde{l}\;
\left\{
\begin{array}{l}
  \phantom{Q}\hspace*{-3mm}\cos\theta\\
  \phantom{Q}\hspace*{-3mm}\sin\theta\,\sin\phi \\
  \phantom{Q}\hspace*{-3mm}\sin\theta\,\cos\phi
\end{array}
 \right\}\,, \;\;\;\;\text{for}\;\; \widetilde{l}\geq \lambda>0\,,
\\[2mm]
\label{eq:Cartesian-coordinates-minus}
\left\{
\begin{array}{c}
  Z_{-} \\
  Y_{-} \\
  X_{-}
\end{array}
 \right\}
&=& \widetilde{l}\;
\left\{
\begin{array}{l}
  \phantom{Q}\hspace*{-3mm}\cos\theta\\
  \phantom{Q}\hspace*{-3mm}\sin\theta\,\sin\phi \\
  \phantom{Q}\hspace*{-3mm}\sin\theta\,\cos\phi
\end{array}
 \right\}\,, \;\;\;\;\text{for}\;\; \widetilde{l}\leq -\lambda<0\,,
\\[2mm]
\label{eq:Cartesian-coordinates-antipodal}
 \left\{Z_{+},\,  Y_{+},\,  X_{+}\right\}
&\stackrel{\wedge}{=}&
 \left\{Z_{-},\,  Y_{-},\,  X_{-}\right\}\,,
 \,\;\;\;\;\;\;\;\;\text{for}\;\; |\,\widetilde{l}\,| = \lambda\,,
\eeqa
\esubeqs
where the last relation implements the identification of
``antipodal'' points on the two \mbox{2-spheres} $S^{\,2}_{\pm}$
with $|\,\widetilde{l}\,|=\lambda$
(the quotation marks are because normally antipodal points
are identified on a \emph{single} 2-sphere,
as for the $\mathbb{R}P^{3}$ defect discussed in
Refs.~\cite{Klinkhamer2014-prd,KlinkhamerSorba2014,Klinkhamer2014-mpla}).
Note that the two coordinate sets  
$\left\{Z_{\pm},\,  Y_{\pm},\,  X_{\pm}\right\}$ 
from \eqref{eq:Cartesian-coordinates-plus}
and \eqref{eq:Cartesian-coordinates-minus} have different orientation
(see the penultimate paragraph of
Sec.~\ref{subsubsec:Vacuum-solution-geodesics} for a further comment).

The spatial topology of our degenerate-wormhole spacetime \eqref{eq:K}
is that of two copies of the Euclidean space $E_3$
with the interior of two balls removed
and ``antipodal'' identification \eqref{eq:Cartesian-coordinates-antipodal}
of their two surfaces. It can be verified that the
wormhole spacetime from \eqref{eq:K} and
\eqref{eq:Cartesian-coordinates} is
simply connected (all loops in space are contractible to a point),
whereas the original exotic-matter wormhole~\cite{MorrisThorne1988}
is  multiply connected
(there are noncontractible loops in space, for example, a
loop in the upper universe encircling the wormhole mouth).

\subsection{Vacuum solution}
\label{subsec:Vacuum-solution} 

\subsubsection{First-order equations}
\label{subsubsec:Vacuum-solution-first-order-equations} 

Awaiting the final analysis of the general metric \eqref{eq:K},
we recall, from Sec.~\ref{subsec:Degenerate-metric-special-case},
that we already have an analytic wormhole-type solution of the
Einstein gravitational field equation
(defined at $\xi=0$ by the limit $\xi \to 0$):
\bsubeqs\label{eq:vacuum-wormhole}
\beqa\label{eq:vacuum-wormhole-metric-functions}
\left\{\widetilde{\phi}(\xi),\,\widetilde{r}^{\;2}(\xi) \right\}\,
\Big|^\text{(K-worm-gen)}_\text{vacuum\;sol}
&=&
\Big\{0,\, \lambda^{2} + \xi^{2}\,\Big\}\,,
\\[4mm]
T^{\,\mu}_{\;\;\,\nu}(\xi)\,
\Big|^\text{(K-worm-gen)}_\text{vacuum\;sol}
&=& 0\,.
\eeqa
\esubeqs
Unlike Minkowski spacetime,
this flat vacuum-wormhole spacetime has asymptotically two
flat 3-spaces with different orientations
(see Sec.~\ref{subsubsec:Vacuum-solution-geodesics} 
for further comments).

Before we turn to the geodesics of the 
vacuum wormhole spacetime \eqref{eq:vacuum-wormhole},
we present an important mathematical result
on the spacetime structure at the wormhole throat, $\xi=0$.
It has been observed by Horowitz~\cite{Horowitz1991}
that the first-order (Palatini) formalism of General Relativity
would be especially suited to the case of degenerate metrics,
the essential point being that the first-order formalism
does not require the inverse metric.
Let us have a look at the degenerate vacuum-wormhole metric
from \eqref{eq:K} and \eqref{eq:vacuum-wormhole}. 
We refer to Refs.~\cite{EguchiGilkeyHanson1980,Nakahara2003} 
for background on Cartan's differential-form approach
and adopt the notation of Ref.~\cite{EguchiGilkeyHanson1980}.

Take, then, the following dual basis 
$e^{a} \equiv e^{a}_{\phantom{z}\mu}\,\text{d}x^{\mu}$\,
from the general expression \eqref{eq:generalized-wormhole-tetrad}
with restrictions \eqref{eq:vacuum-wormhole-metric-functions}:
\bsubeqs\label{eq:vacuum-wormhole-tetrad}
\beqa
e^{0}\,\Big|^\text{(K-worm-gen)}_\text{vacuum\;sol}
&=& \text{d}t\,,
\\[2mm]
\label{eq:vacuum-wormhole-tetrad-a-is-1}
e^{1}\,\Big|^\text{(K-worm-gen)}_\text{vacuum\;sol}
&=&  \frac{\xi}{\sqrt{\lambda^{2} + \xi^{2}}}\; \text{d}\xi\,,
\\[2mm]
e^{2}\,\Big|^\text{(K-worm-gen)}_\text{vacuum\;sol}
&=&  \sqrt{\lambda^{2} + \xi^{2}}\; \text{d}\theta\,,
\\[2mm]
e^{3}\,\Big|^\text{(K-worm-gen)}_\text{vacuum\;sol}
&=& \sqrt{\lambda^{2} + \xi^{2}}\;\sin\theta  \;  \text{d}\phi\,.
\eeqa
\esubeqs
This basis gives, from the metricity condition 
($\omega_{ab}=-\omega_{ba}$) and the no-torsion 
condition ($\text{d}e^{a}+\omega^{\,a}_{\,\phantom{z}b}\wedge e^{b}=0$), 
the following nonzero components of the Levi--Civita spin connection:
\beq
\label{eq:vacuum-wormhole-connection}
\left\{
\omega^{2}_{\phantom{z}1},\,  
\omega^{3}_{\phantom{z}1},\,
\omega^{3}_{\phantom{z}2}
\right\}\,\Big|^\text{(K-worm-gen)}_\text{vacuum\;sol}
=
\left\{
\text{d}\theta,\, 
\sin\theta  \;  \text{d}\phi,\,
\cos\theta  \;  \text{d}\phi
\right\}\,,
\eeq
with corresponding components 
$\left\{\omega^{1}_{\phantom{z}2},\,  
\omega^{1}_{\phantom{z}3},\,\omega^{2}_{\phantom{z}3}\right\}$
by antisymmetry.
The resulting curvature 2-form 
$R^{\,a}_{\,\phantom{z}b} \equiv \text{d}\omega^{\,a}_{\,\phantom{z}b}+
\omega^{a}_{\phantom{z}c}  \wedge \omega^{c}_{\phantom{z}b}$
has all components vanishing identically,
\beq
\label{eq:vacuum-wormhole-curvature}
R^{\,a}_{\phantom{z}b}\,\Big|^\text{(K-worm-gen)}_\text{vacuum\;sol}
=0\,.
\eeq
The crucial observation is that the above spin-connection
and curvature components are well-behaved at $\xi=0$, so that
there is no direct need for the $\xi \to 0$ limit.

All in all, the degenerate vacuum-wormhole metric 
from \eqref{eq:K} and \eqref{eq:vacuum-wormhole}
provides a smooth solution 
\eqref{eq:vacuum-wormhole-tetrad}--\eqref{eq:vacuum-wormhole-connection}
of the first-order equations of General Relativity~\cite{Horowitz1991},
\bsubeqs\label{eq:first-order-eqs} 
\beqa\label{eq:first-order-eqs-no-torsion}
e^{\,[\,a} \wedge D\, e^{\,b\,]} &=& 0\,,
\\[2mm]
\label{eq:first-order-eqs-Ricci-flat}
e^{\,b} \wedge R^{\,cd}\,\epsilon_{abcd} &=& 0 \,,
\eeqa
\esubeqs
with the completely antisymmetric symbol $\epsilon_{abcd}$,
the covariant derivative
$D\, e^{b} \equiv \text{d}e^{b}+\omega^{\,b}_{\,\phantom{z}c}\wedge e^{c}$,
and the square brackets around Lorentz indices $a$ and $b$ denoting
antisymmetrization.
We see that the complete vacuum solution is given by the  
tetrad $e^{a}_{\mu}(x)$ from \eqref{eq:vacuum-wormhole-tetrad}
and the connection $\omega^{\phantom{z}a}_{\mu\phantom{z}b}(x)$  
from \eqref{eq:vacuum-wormhole-connection},
not just the metric $g_{\mu\nu}(x)$
from \eqref{eq:K} and \eqref{eq:vacuum-wormhole}.

\subsubsection{Geodesics}
\label{subsubsec:Vacuum-solution-geodesics} 

We now get explicitly the radial geodesics $\xi(t)$
passing through the vacuum-wormhole throat by adapting
result (3.6b) of Ref.~\cite{Klinkhamer2014-mpla} to our case:
\beqa\label{eq:radial-geodesics}
\xi(t)\,\Big|^\text{(K-worm-gen)}_\text{vacuum\;sol\,;\;rad-geod}
&=&
\begin{cases}
 \pm\,\sqrt{(B\,t)^{2}+2\,B\,\lambda\,t} \,,   &  \;\;\text{for}\;\; t \geq 0 \,,
 \\[2mm]
 \mp\,\sqrt{(B\,t)^{2}-2\,B\,\lambda\,t} \,,   &  \;\;\text{for}\;\; t \leq 0 \,,
\end{cases}
\eeqa
with a dimensionless constant $B \in (0,\,1]$
and different signs (upper or lower)
in front of the square roots for motion in opposite directions.
The same curves \eqref{eq:radial-geodesics} can be more
easily obtained from straight lines $\widetilde{l}(t)$,
with radial velocity magnitude $v$,
in the 2-dimensional Minkowski subspace
of the spacetime \eqref{eq:ltilde-metric}
and the definition \eqref{eq:ltilde}
of the coordinate $\widetilde{l}$.
The Minkowski-subspace analysis identifies the
constant $B$ in \eqref{eq:radial-geodesics} with the
ratio $v/c$, so that the $B=1$ curves are light-like and
those with $B<1$ timelike. For a more detailed description of
these geodesics, it appears worthwhile to change
to the Cartesian coordinates of Sec.~\ref{subsec:Topology}.

Consider, indeed, the radial geodesic \eqref{eq:radial-geodesics}
with the upper signs and fixed $\{\theta ,\, \phi \}= \{\pi/2 ,\, 0 \}$
and obtain the trajectory in terms of the
Cartesian coordinates \eqref{eq:Cartesian-coordinates}:
\bsubeqs\label{eq:radial-geodesic-Cartesian-coord}
\beqa
Z_{\pm}(t)\,\Big|^\text{(K-worm-gen)}_\text{vacuum\;sol\,;\;rad-geod}
&=&  0\,,\qquad\qquad\;\; \text{for}\;\; t \in (-\infty,\,\infty)\,,
\\[2mm]
Y_{\pm}(t)\,\Big|^\text{(K-worm-gen)}_\text{vacuum\;sol\,;\;rad-geod}
&=&  0\,,\qquad\qquad\;\; \text{for}\;\; t \in (-\infty,\,\infty)\,,
\\[2mm]
X_{-}(t)\,\Big|^\text{(K-worm-gen)}_\text{vacuum\;sol\,;\;rad-geod}
&=&
 -\lambda+B\,t\,,\;\;\;\;\text{for}\;\; t \leq 0 \,,
\\[2mm]
X_{+}(t)\,\Big|^\text{(K-worm-gen)}_\text{vacuum\;sol\,;\;rad-geod}
&=& +\lambda+B\,t\,,\;\;\;\;\text{for}\;\; t \ge 0 \,,
\eeqa
\esubeqs
with $X_{-}=-\lambda$ and $X_{+}=+\lambda$ identified at $t=0$.
The apparent discontinuity of $X_{\pm}$ is an artifact of using
two copies of Euclidean 3-space for the embedding of the
trajectory, but the curve on the real manifold is smooth,
as shown by \eqref{eq:radial-geodesics}. This point is
also illustrated for the $\mathbb{R}P^{3}$ defect by Figs.~2 and 3
in Ref.~\cite{Klinkhamer2014-mpla}.

The curves from \eqref{eq:radial-geodesic-Cartesian-coord}
in the  $(t,\,X_{-})$ and $(t,\,X_{+})$ planes,
have two parallel straight-line segments, shifted at $t=0$,
with constant positive slope $B\leq 1$ (velocity in units with $c=1$).
This equal velocity before and after the defect crossing
is the main argument for using the ``antipodal''
identification in \eqref{eq:Cartesian-coordinates}.
This identification also agrees with the observation
that the tetrad from \eqref{eq:vacuum-wormhole-tetrad} provides
a smooth solution 
of the first-order equations of General Relativity, 
which would not be the case for the tetrad  
with $\xi$ in the numerator on the right-hand side of
\eqref{eq:vacuum-wormhole-tetrad-a-is-1} replaced by $\sqrt{\xi^{2}}$.

The discussion of nonradial geodesics for the
metric \eqref{eq:vacuum-wormhole} is similar to the
discussion in Ref.~\cite{KlinkhamerWang2019},
which considers a related spacetime defect.
The metric of this spacetime defect resembles
the metric of the wormhole presented here, but
their global spatial structures are different.
Still, it appears possible to take over the
results from Ref.~\cite{KlinkhamerWang2019} on
defect-crossing nonradial geodesics (which stay in
a single universe), if we realize that, for our vacuum wormhole,
the defect-crossing geodesics come in from
one universe and re-emerge in the other universe.
These nonradial geodesics of the vacuum wormhole  
will be discussed later.

Following up on the issue of spatial orientability
mentioned in the first paragraph of 
Sec.~\ref{subsubsec:Vacuum-solution-first-order-equations},
we have the following comment.
If the ``advanced civilization'' of Ref.~\cite{MorrisThorne1988}
has access to our type of defect-wormhole, then it
is perhaps advisable to start exploration by sending in
parity-invariant machines or robots. The reason is that
humans of finite size and with right-handed DNA may not be able
to pass safely through this particular wormhole-throat defect
at $\xi=0$,
which separates two universes with different 3-space orientation.

The vacuum solution \eqref{eq:vacuum-wormhole} 
with tetrad \eqref{eq:vacuum-wormhole-tetrad} 
and connection \eqref{eq:vacuum-wormhole-connection} 
has one free parameter,
the length scale $\lambda$ which can, in principle, be determined
as the limiting value of circumferences divided by $2\pi$ for
great circles centered on the wormhole mouth
[in the metric \eqref{eq:ltilde-metric} for $\lambda^{2}=b_{0}^{2}\,$,
circles with, for example, $\phi \in [0,\,2\pi)$, constant $\theta=\pi/2$,
constant $t$, and various positive values of $\widetilde{l}$ 
(keeping the same spatial orientation)].
An alternative way to determine the length scale $\lambda$  
of a single vacuum-defect wormhole
is to measure its lensing effects
(cf. Sec.~5 of Ref.~\cite{KlinkhamerWang2019})
and an explicit example is presented in
App.~\ref{app:Gedankenexperiment}.
Some further comments on this length scale $\lambda$
appear in Sec.~\ref{sec:Discussion}.

\section{Discussion}
\label{sec:Discussion}

We have five final remarks. 
First, we have obtained, in line with earlier
work by Horowitz~\cite{Horowitz1991},
a smooth vacuum-wormhole solution of the first-order equations 
of General Relativity,
where the tetrad is given by \eqref{eq:vacuum-wormhole-tetrad} 
and the connection by \eqref{eq:vacuum-wormhole-connection}. 
Vacuum wormholes also appear in certain modified-gravity
theories (see, e.g., Ref.~\cite{CalzaRinaldiSebastiani2018}
and references therein), where the
exotic effects trace back to the extra terms in the
gravitational action (see also  
Refs.~\cite{KarLahiriSenGupta-2015,KonoplyaZhidenko2022,%
Terno2022,Biswas-etal2022} and references therein).
Our vacuum-wormhole solution
does not require any fundamental change of the theory,
4-dimensional General Relativity suffices,
except that we now allow for degenerate metrics.
The degeneracy hypersurface of the vacuum-wormhole solution
corresponds to a ``spacetime defect,'' as discussed extensively in
Refs.~\cite{Klinkhamer2019a,Klinkhamer2019b,KlinkhamerWang2019a,KlinkhamerWang2019b,
Klinkhamer2021-review,Klinkhamer2014-prd,KlinkhamerSorba2014,%
Guenther2017,Klinkhamer2019-jpcs,Klinkhamer2014-mpla,KlinkhamerWang2019}.

Second, this vacuum-defect-wormhole solution \eqref{eq:vacuum-wormhole-tetrad}  
has the length scale $\lambda$ as a free parameter and, if there
is a preferred value $\overline{\lambda}$ in Nature, then that value
can only come from a theory beyond General Relativity.
An example of such a theory would be  nonperturbative superstring theory
in the formulation of the IIB matrix
model~\cite{IKKT-1997,Aoki-etal-review-1999}.
That matrix model could give rise to an emergent
spacetime with or without spacetime
defects~\cite{Klinkhamer2021-master,Klinkhamer2021-regBB}
and, if defects do appear,
then the typical length scale $\overline{\lambda}$ of a remnant
vacuum-wormhole defect would be related
to the IIB-matrix-model length scale $\ell$
(the Planck length $G^{1/2}$ might
also be related to this length scale $\ell$\,).

Third, the main objective of the present paper
has been to reduce the hurdles to overcome
in the quest of traversable wormholes
(specifically, we have removed the requirement of exotic matter).
But there remains, at least, one important hurdle,
namely to construct a suitable spacetime defect
or to harvest one, if already present as a remnant
from an early phase.

Fourth, if it is indeed possible to harvest a vacuum-defect wormhole,
then its length scale $\lambda$ is most likely very small
(perhaps of the order of
the Planck length $G^{1/2} \sim 10^{-35}\,\text{m}$).
The question now arises if it is,
in principle, feasible to enlarge (fatten) that harvested
defect wormhole. Preliminary numerical results
presented in App.~\ref{app:Preliminary-numerical-results}
suggest that the answer may be affirmative.

Fifth, the construction of multiple 
vacuum-defect-wormhole solutions is relatively straightforward,  
provided the respective wormhole throats do not 
touch. Details and further discussion are
given in a follow-up paper~\cite{Klinkhamer2023-mirror-world}.

\begin{acknowledgments}
It is a pleasure to thank Z.L.~Wang for useful comments on the manuscript
and E.~Guendelman for a helpful remark after a recent 
wormhole talk by the author. The referee is thanked for a
practical suggestion to improve the presentation.
\end{acknowledgments}

\begin{appendix}

\section{Gedankenexperiment}
\label{app:Gedankenexperiment}

\begin{figure}[t]   
\vspace*{-2mm}  
\begin{center}
\includegraphics[width=0.45\textwidth]{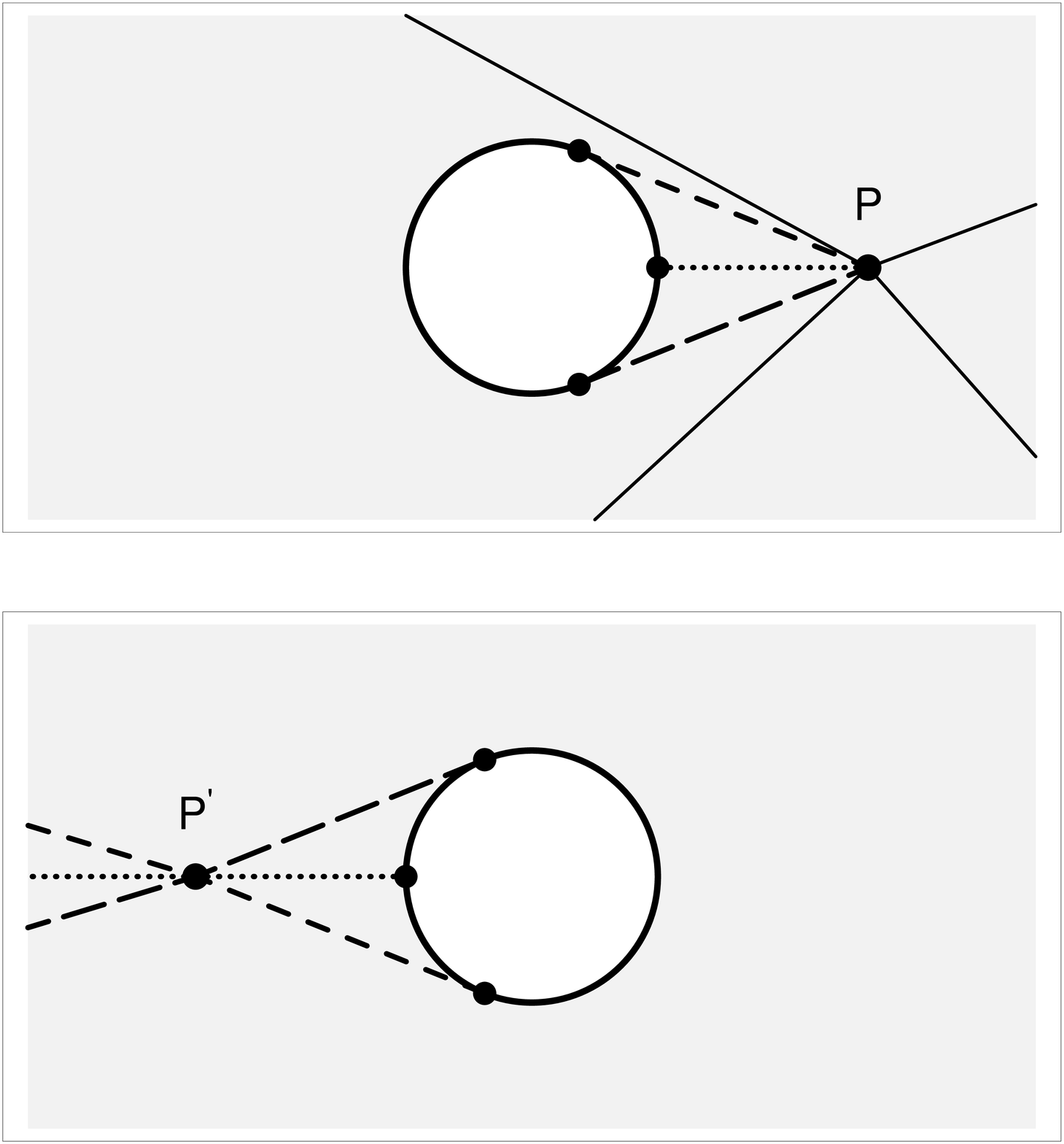}
\end{center}
\vspace*{-5mm}
\caption{Geodesics emanating from point $\text{P}$
 in the ``upper'' universe of
the vacuum-defect-wormhole spacetime with 
metric \eqref{eq:K} and \eqref{eq:vacuum-wormhole}. 
Certain geodesics (thin solid lines)
avoid the wormhole throat and stay in the upper universe,
while other geodesics (heavy short-dashed/long-dashed/dotted lines)
reach the wormhole throat (here shown as a thick circle
with antipodal points identified)
and cross over into the ``lower'' universe.
The wormhole acts as a lens, with the transmitted geodesics refocussing
at the point $\text{P}^{\,\prime}$ in the ``lower'' universe.
}
\label{fig:Lensing}
\vspace*{-2mm}
\end{figure}

In this appendix, we describe a \textit{Gedankenexperiment}
designed to 
measure the length scale $\lambda$ of a vacuum-defect wormhole
by its lensing effects.
The lensing property is illustrated in Fig.~\ref{fig:Lensing}.
More specifically, the \textit{Gedankenexperiment}
is based on the fifth remark of Sec.~5 in Ref.~\cite{KlinkhamerWang2019},
which, adapted to our case, states that 
a permanent point-like source at point $\text{P}$ 
from Fig.~\ref{fig:Lensing} 
will be seen as a luminous disk at point $\text{P}^{\,\prime}$
from the same figure.
(The lensing properties have, of course, first been
discussed for exotic-matter wormholes and 
a selection of references appears in 
Ref.~\cite{ChetouaniClement1984,Perlick2004,%
Nandi-etal2006,TsukamotoHarada2016,Shaikh-etal-2018}.)

The concrete procedure of the \textit{Gedankenexperiment}
involves three steps (cf. Fig.~\ref{fig:Lensing}):
\begin{enumerate}
  \item 
place a \emph{permanent point-like light source} 
at an arbitrary point $\text{P}$, 
\item 
search for the point $\text{P}^{\,\prime}$ where 
a \emph{luminous disk} is seen;
\item 
\emph{measure} two quantities, the angle $2\,\alpha$ that the
disk subtends as seen from $\text{P}^{\,\prime}$ and the shortest
distance $D_{\,\text{PP}'}$ between the points $\text{P}$ and $\text{P}^{\,\prime}$.
\end{enumerate}
If no point with a luminous disk can be found  in Step 2, then
the point $\text{P}$ must have been on the wormhole throat and we
return to Step 1 by changing the position of point $\text{P}$.

Now use the auxiliary coordinates and metric 
from \eqref{eq:ltilde-metric} 
for $ b_{0}^{2}=\lambda^{2}$.
With $\widetilde{l}_\text{\,P} > \lambda$ denoting the
quasi-radial coordinate of point $\text{P}$ (assumed
to be in the  ``upper'' universe), 
we have $D_{\,\text{PP}'}=2\,\big(\,\widetilde{l}_\text{\,P}-\lambda\big)$
and $\sin\alpha= \lambda/\,\widetilde{l}_\text{\,P}$.  
Combined, these two expressions give
the following result for 
the length scale $\lambda$ of the vacuum-defect wormhole:
\beq
\label{eq:lambda}  
\lambda=
\frac{\sin\alpha}{1-\sin\alpha}\;\;\frac{D_{\,\text{PP}'}}{2}\,,
\eeq
solely in terms of the measured quantities $\alpha$ and $D_{\,\text{PP}'}$.

We have two further observations. First, recall that 
a permanent point-like source in Minkowski spacetime will,
in principle, illuminate the whole of 3-space. But, in
the vacuum-defect-wormhole spacetime, 
there will exist dark regions, 
even in the upper universe where the source is located.
In the upper universe of Fig.~\ref{fig:Lensing}, 
there is indeed a dark (shadow) region 
behind the wormhole throat (which has drained 
away some of the light emitted by the source). 
It is a straightforward exercise to describe 
the dark regions exactly.

Second, if the light source at $\text{P}$ is no longer permanent
but a flash instead,
then we observe at $\text{P}^{\,\prime}$, after a
certain moment, an expanding ring,
even though there is no motion of the source position.
This effect is entirely due to differences in the time-of-flight,
for example, the time-of-flight of the short-dashed 
geodesic in Fig.~\ref{fig:Lensing} is larger than that of the dotted geodesic. 
The maximal angular extension ($2\alpha$) of the expanding ring
and the source-observer distance ($D_{\,\text{PP}'}$) can, in principle,
give the vacuum-defect-wormhole length scale $\lambda$
by the expression \eqref{eq:lambda}.

\section{Preliminary numerical results}
\label{app:Preliminary-numerical-results}

In the general \textit{Ansatz} \eqref{eq:K}
for the defect-wormhole  metric, we set
\bsubeqs\label{eq:Ansaetze-phitilde-rtilde2}
\beqa
\label{eq:Ansatz-phitilde}
\widetilde{\phi}(\xi)&=& f(\xi)\,,
\\[2mm]
\label{eq:Ansatz-rtilde2}
\widetilde{r}^{\;2}(\xi)&=&
\lambda^{2}+\xi^{2} +g(\xi)\,,
\eeqa
\esubeqs
so that a nonzero function $g(\xi)$
signals a non-vacuum wormhole configuration.
For the numerical analysis, it turns out to be useful
to compactify the quasi-radial $\xi$ coordinate,
\beq
\label{eq:eta}
\eta = \text{sgn}(\xi)\;\frac{\xi^{2}}{\lambda^{2}+\xi^{2}}
\in [-1,\,1]\,.
\eeq
In the following, we consider only nonnegative $\eta$ and $\xi$.

Next, we expand the metric functions
$\overline{f}(\eta)=f(\xi)$ and $\overline{g}(\eta)=g(\xi)$
over $\eta \in [0,\,1]$,
\bsubeqs\label{eq:expansions-fbar-gbar}
\beqa
\overline{f}(\eta) &=& \sum_{n=1}^{N_{f}}\,c_{n}\,\sin\big(n\,\pi\,\eta\big)\,,
\\[2mm]
\overline{g}(\eta) &=&
d_{0}\,(1-\eta)+\sum_{n=1}^{N_{g}}\,d_{n}\,\sin\big(n\,\pi\,\eta\big)\,,
\eeqa
\esubeqs
with finite cutoffs $N_{f}$ and $N_{g}$ on the sums.
In this appendix, we take
\beq
\label{eq:lambdanum-d0num}
\lambda=1 \,, \quad d_{0}=1/10 \,,
\eeq
where the small but nonzero ratio $d_{0}/\lambda^{2}$
quantifies the deviation of the vacuum configuration.

Consider the radial null vector
$\widehat{k}_{\,\mu}$ $=$
$\big(\exp\big[\,\overline{f}(\eta)\big],\,\sqrt{|\eta|},\,0,\,0\big)$
and the tangential null vector
$\overline{k}_{\,\mu}$ $=$
$\big(\exp\big[\,\overline{f}(\eta)\big],\,0,\,\sqrt{\widetilde{r}^{\:2}},\,0\big)$,  
with replacement \eqref{eq:Ansatz-rtilde2} in terms of $\eta$.
Then define the following quantities:
\bsubeqs\label{eq:Tkkrad-Tkktang}
\beqa
\Theta_\text{rad}
\equiv
G^{\,\mu}_{\;\;\,\nu}\,\widehat{k}_{\mu}\,\widehat{k}^{\,\nu}\,,
\\[2mm]
\Theta_\text{tang}
\equiv
G^{\,\mu}_{\;\;\,\nu}\,\overline{k}_{\mu}\,\overline{k}^{\,\nu}\,,
\eeqa
\esubeqs
where $G^{\,\mu}_{\;\;\,\nu}\equiv
R^{\,\mu}_{\;\;\,\nu} - \half\, g^{\,\mu}_{\;\;\,\nu}\,R$
is the Einstein tensor,
which equals the energy-momentum tensor $T^{\,\mu}_{\;\;\,\nu}$
from the assumed validity of the Einstein equation for
units with $8\pi G=1$.

We, now, introduce the following ``penalty'' measure:
\beq
\label{eq:pen}
\mathcal{P}\equiv
\int_{0}^{1}\, d\eta \,
\left(
\Bigg[ \sqrt{(\Theta_\text{rad})^{2}}  - \Theta_\text{rad} \,\Bigg]^{2} +
\Bigg[ \sqrt{(\Theta_\text{tang})^{2}} - \Theta_\text{tang}\,\Bigg]^{2}\,
\right) \,.
\eeq
The Null Energy Condition (NEC), as discussed in
Sec.~\ref{sec:Basic-idea}, gives $\mathcal{P}=0$.
With the \textit{ad hoc} expansion \eqref{eq:expansions-fbar-gbar},
the quantity $\mathcal{P}$ from
\eqref{eq:pen} is a function of the coefficients
$c_{n}$ and $d_{n}$, which can be minimized numerically.
We have used
the \texttt{NMinimize} routine of \textsc{Mathematica} 5.0.  

\begin{figure}[t]
\vspace*{-0mm}  
\begin{center}
\includegraphics[width=0.75\textwidth]{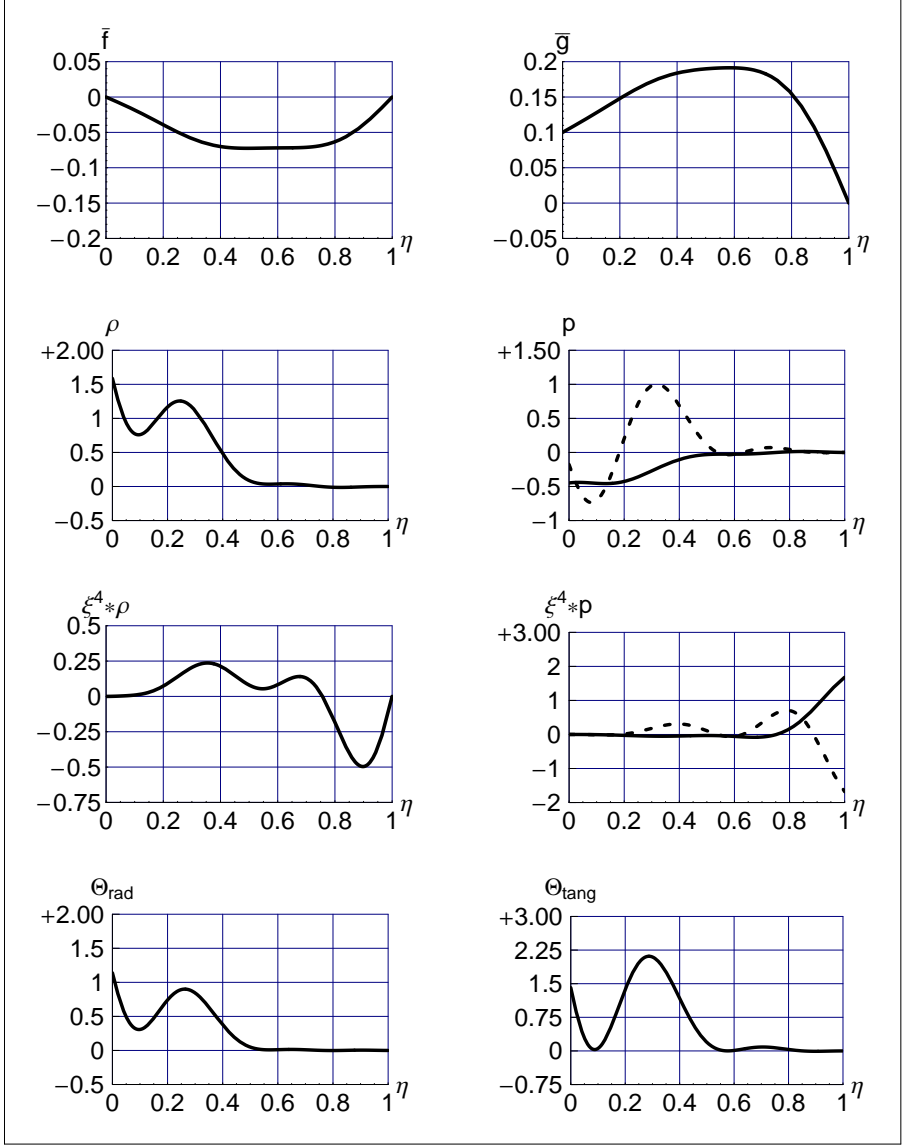}
\end{center}
\vspace*{-0mm}
\caption{Metric \textit{Ansatz} functions
$\overline{f}(\eta)=f(\xi)$ and $\overline{g}(\eta)=g(\xi)$
in the metric \eqref{eq:K} as defined
by \eqref{eq:Ansaetze-phitilde-rtilde2}, \eqref{eq:eta},
and \eqref{eq:expansions-fbar-gbar}
for the eight coefficients from Table~\ref{tab:Eight-coeff},
giving for the penalty function $\mathcal{P}$  
from \eqref{eq:pen} the value $1.72084 \times 10^{-5}$.
In addition, there are the following numerical
parameters: $\lambda=1$ and $d_{0}=1/10$.
Several quantities result from the basic functions
$\overline{f}(\eta)$ and $\overline{g}(\eta)$.
The second row shows, at the left, the energy density
$\rho \equiv G^{\,t}_{\;\;\,t}$ and, at the right, the pressures
$p_\text{rad} \equiv G^{\,\xi}_{\;\;\,\xi}$ [solid curve]
and $p_\text{tang} \equiv G^{\,\theta}_{\;\;\,\theta}$ [dashed curve],
where $G^{\,\mu}_{\;\;\,\nu}$ is the Einstein tensor.
On the third row, these last three quantities have been multiplied by
$\xi^4 = [\lambda^{2}\,\eta/(1-\eta)]^{2}$, in order to display
their rapid asymptotic decrease
(approximately as $\xi^{-4}$ for $\xi \to \infty$  or $\eta \to 1$).
The fourth row shows the quantities
$\Theta_\text{rad}$ and $\Theta_\text{tang}$
as defined by \eqref{eq:Tkkrad-Tkktang}.
The Null Energy Condition would correspond to having
$\Theta_\text{rad} \geq 0$ and
$\Theta_\text{tang}  \geq 0$. With only eight coefficients,
there are still small Null-Energy-Condition violations
of $\Theta_\text{rad}$ at $\eta \sim 0.8$
and $\Theta_\text{tang}$ at $\eta \sim 0.9$;
see the main text in
App.~\ref{app:Preliminary-numerical-results} for further discussion.}
\label{fig:all-results-Ncoeff-8}
\vspace*{100mm}
\end{figure}

\begin{figure}[t] 
\vspace*{-0mm}    
\begin{center}
\includegraphics[width=0.75\textwidth]{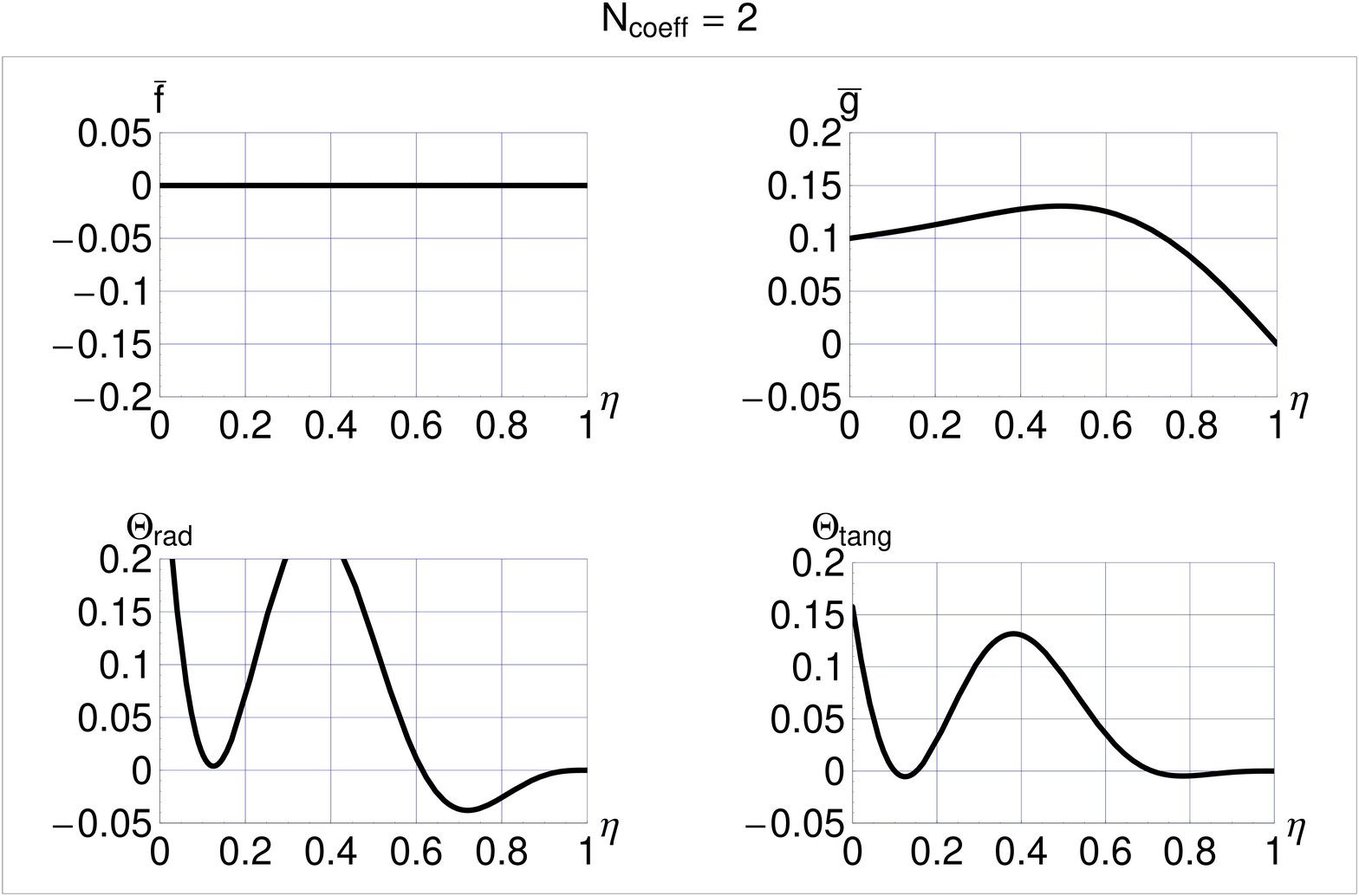}
\end{center}
\vspace*{-5mm}
\vspace*{4mm}
\vspace*{-0mm}   
\begin{center}   
\includegraphics[width=0.75\textwidth]{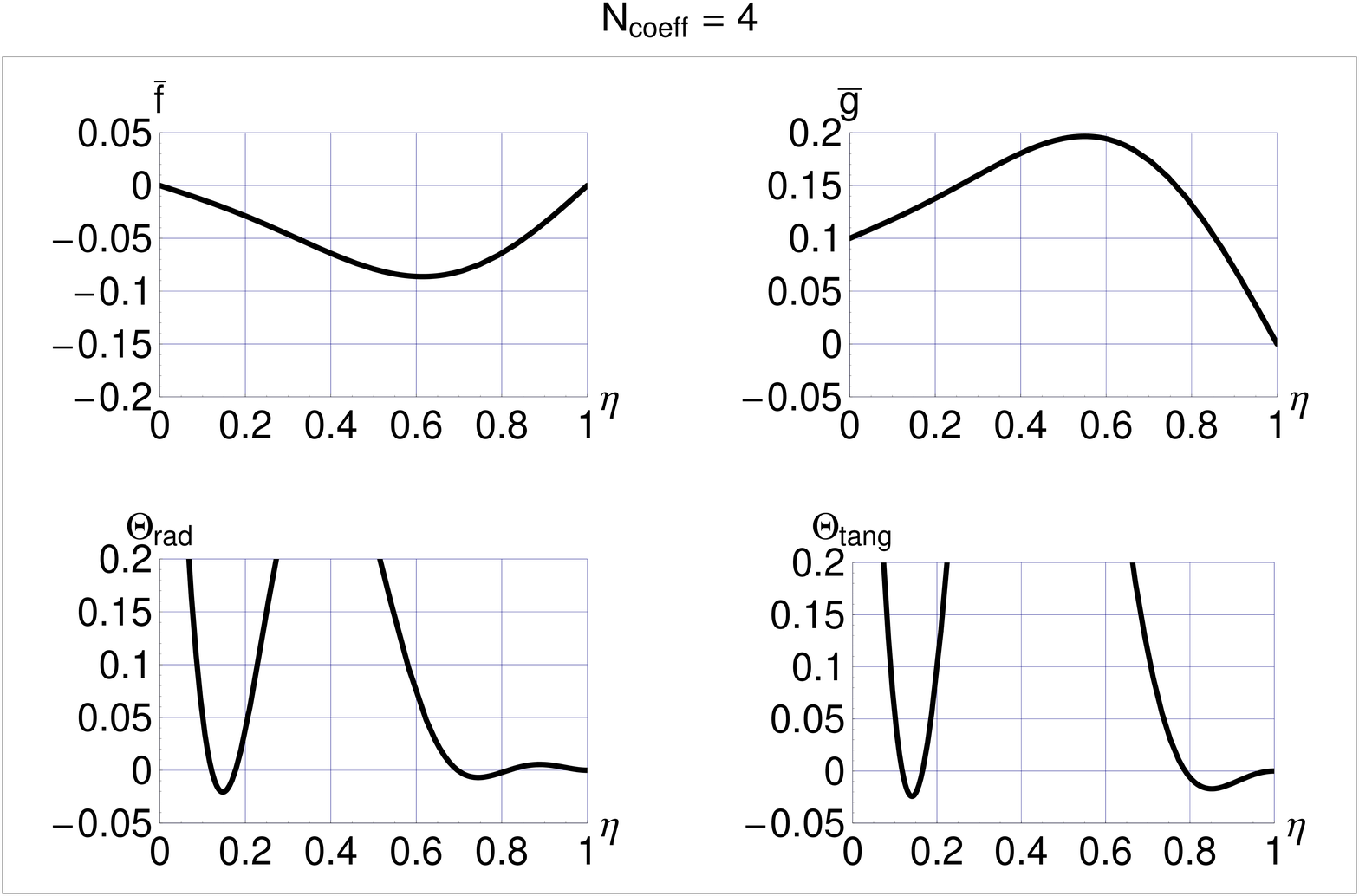}
\end{center}
\vspace*{-0mm}
\caption{Numerical results for $N_\text{coeff} \equiv N_{f} + N_{g} =2$
in the top panel 
[penalty $\mathcal{P}=0.000812553$
for nonzero coefficients $d_{1}=0.0805092$ and $d_{2}=-0.0151493$]
and for $N_\text{coeff}=4$ in the bottom 
panel 
[penalty $\mathcal{P}=0.000111302$ for nonzero coefficients
$c_{1}=-0.0789967$, $c_{2}=0.0186060$, $d_{1}=0.144524$, and
$d_{2}=-0.0286358$].
}
\label{fig:trend-Ncoeff-2and4}
\vspace*{100mm}
\end{figure}

\begin{figure}[t]
\vspace*{-0mm}   
\begin{center}
\includegraphics[width=0.75\textwidth]{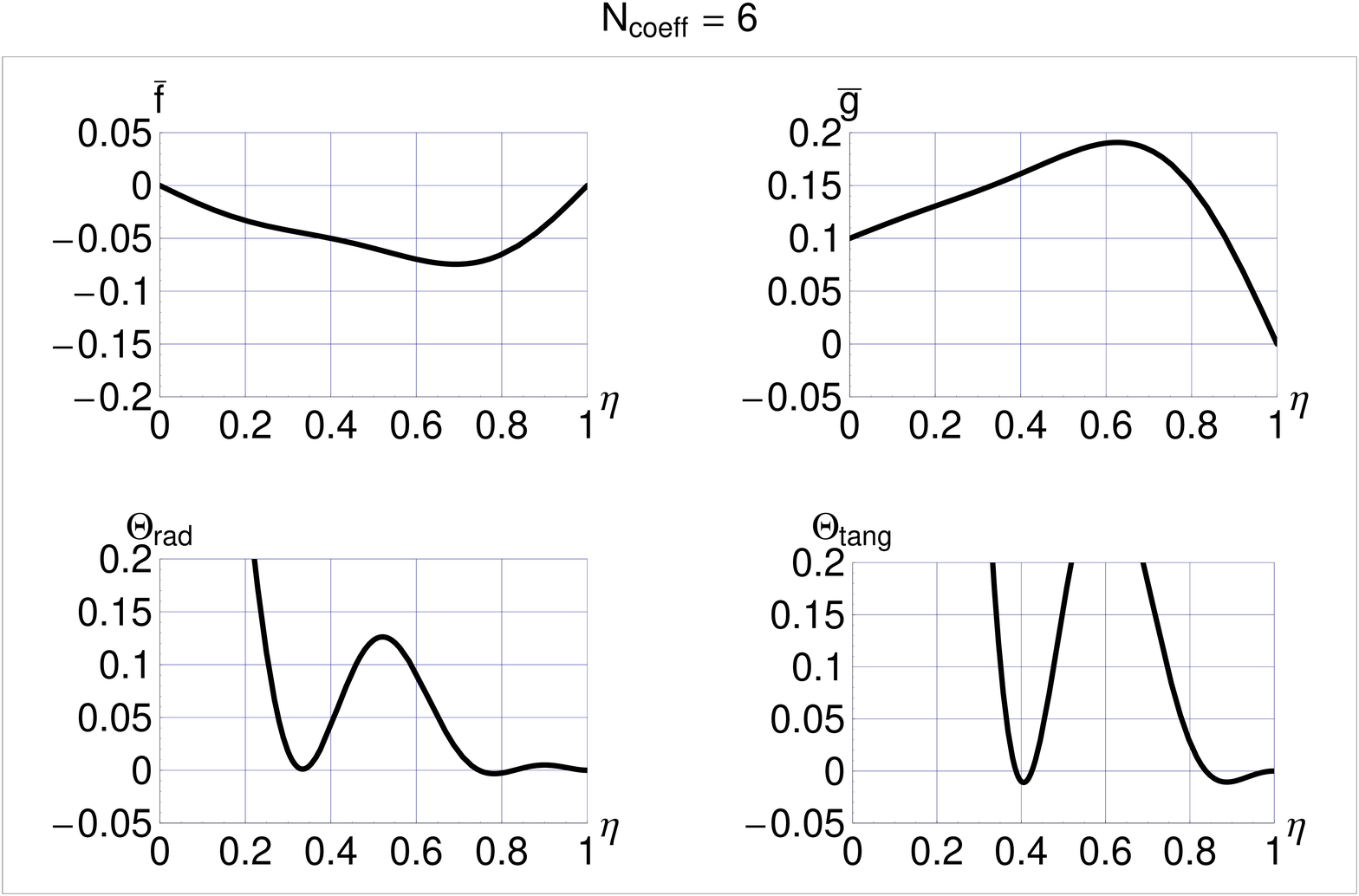}
\end{center}
\vspace*{-5mm}
\vspace*{4mm}
\vspace*{-0mm}  
\begin{center}
\includegraphics[width=0.75\textwidth]{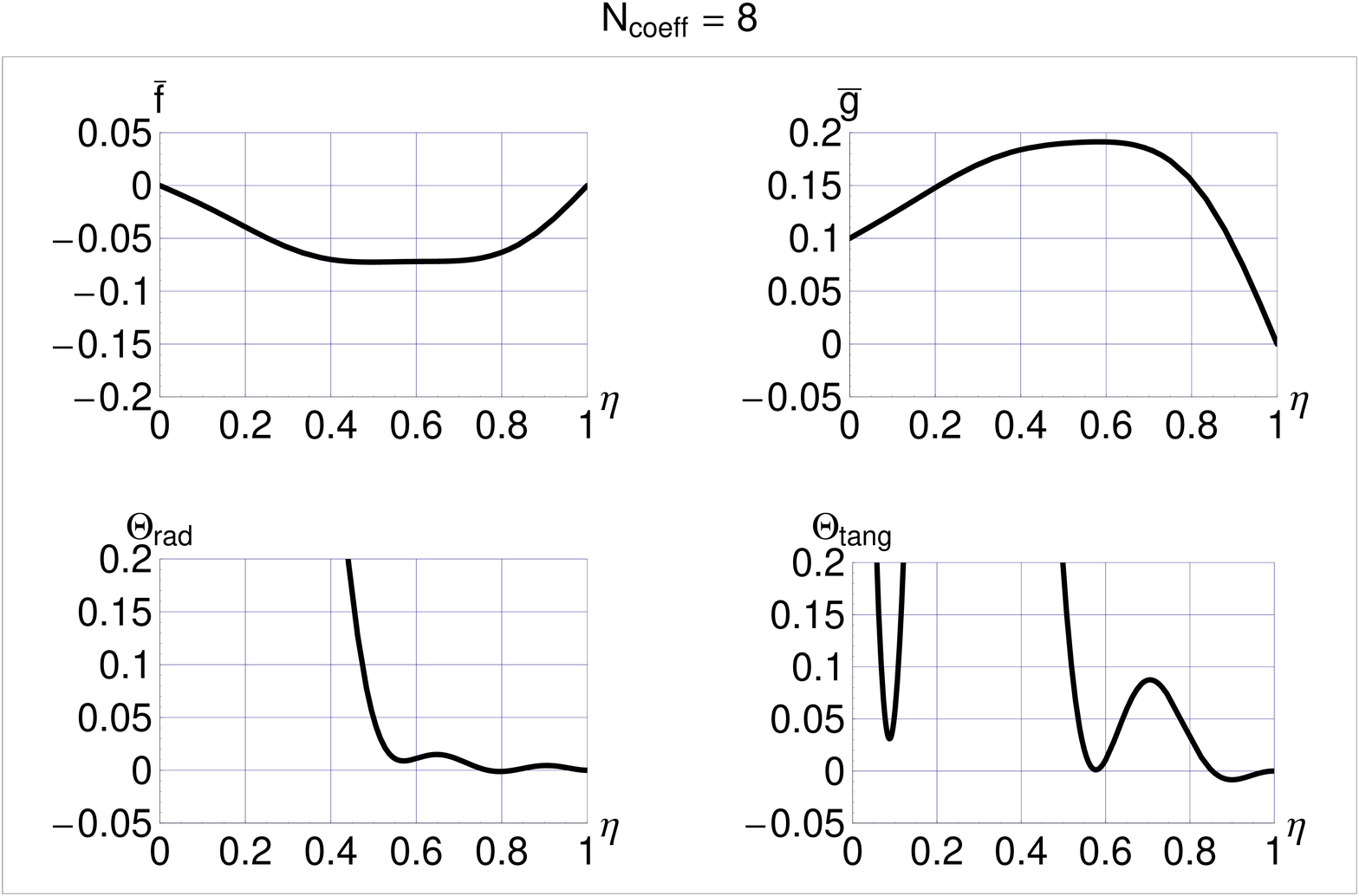}
\end{center}
\vspace*{-0mm}
\caption{Numerical results for $N_\text{coeff}=6$
in the top 
panel  
[penalty $\mathcal{P}=0.0000316721$ for nonzero coefficients
$c_{1}=-0.0686991$, $c_{2}=0.0168485$, $c_{3}=-0.00924705$,
$d_{1}=0.137967$, $d_{2}=-0.0416794$, and $d_{3}=0.00952128$]
and for $N_\text{coeff}=8$ in the bottom 
panel 
[penalty $\mathcal{P}=0.0000172084$ for
the nonzero coefficients from Table~\ref{tab:Eight-coeff}\,].
For $N_\text{coeff}=8$, the $\Theta_\text{rad}$ value  
drops to approximately $-0.001$ at $\eta \sim 0.8$
and the $\Theta_\text{tang}$ value to approximately $-0.01$ 
at $\eta \sim 0.9$.
}
\label{fig:trend-Ncoeff-6and8}
\vspace*{100mm}
\end{figure}

With the eight coefficients from Table~\ref{tab:Eight-coeff},
we are able to reduce the
penalty $\mathcal{P}$ to a value of order $10^{-5}$ and the corresponding
results are shown in Fig.~\ref{fig:all-results-Ncoeff-8}.
Even more important than this small number for $\mathcal{P}$
is the fact that we have established a
\emph{trend} of dropping $\mathcal{P}$ for
an increasing number of coefficients, as shown in
Figs.~\ref{fig:trend-Ncoeff-2and4} and \ref{fig:trend-Ncoeff-6and8}.
Observe also that the function shapes of
$\overline{f}(\eta)$ and $\overline{g}(\eta)$
for $N_\text{coeff} \equiv N_{f} + N_{g} =4,\,6,\,8$
in Figs.~\ref{fig:trend-Ncoeff-2and4} and \ref{fig:trend-Ncoeff-6and8}
are more or less stable and that the absolute values of the
the coefficients $c_{n}$ and $d_{n}$ 
in Table~\ref{tab:Eight-coeff} drop for increasing order $n$. 
All this suggests that the numerical results
converge on a nontrivial configuration,
but this needs to be established rigorously.

\begin{table}[t]
\caption{\label{tab:Eight-coeff}
Eight nonzero coefficients giving $\mathcal{P} = 1.72084 \times 10^{-5}$,
for numerical parameters $\lambda=1$ and $d_{0}=1/10$.  
The coefficients shown are exact numbers and can also be
written as rational numbers, for example
$d_{1}=152362/10^{6}$.\vspace*{3mm}}
\begin{ruledtabular}
\renewcommand{\tabcolsep}{2.5pc}  
\renewcommand{\arraystretch}{1.}  
\begin{tabular}{ll}
$\hspace*{10mm}c_{1}=-0.0781046$ &  $d_{1}=\phantom{+}0.152362\hspace*{10mm}$\\
$\hspace*{10mm}c_{2}=\phantom{+}0.00959397$ &  $d_{2}=-0.0319350\hspace*{10mm}$\\
$\hspace*{10mm}c_{3}=-0.00560390$ &  $d_{3}=\phantom{+}0.0125051\hspace*{10mm}$\\
$\hspace*{10mm}c_{4}= \phantom{+}0.00491741$ &  $d_{4}=-0.00533132\hspace*{10mm}$\\
\end{tabular}
\end{ruledtabular}
\end{table}

The numerical results as they stand appear to indicate that
\begin{enumerate}
\item
the obtained wormhole throat is larger than the one of the
original (harvested) vacuum-defect wormhole,
$\min_{\xi}\big[\widetilde{r}^{\;2}(\xi)\big] > \lambda^{2}$\,;
\item
the energy density $\rho$
and the pressures $\{p_\text{rad} ,\, p_\text{tang}\}$
go to zero as $\xi^{-4}$ asymptotically;
\item 
there is only a small violation of the NEC far away from
the wormhole throat [this NEC violation
can be expected to vanish for an infinite number
of appropriate coefficients 
and the resulting energy density $\rho(\eta)$ is perhaps 
nonnegative everywhere].
\end{enumerate}
If these preliminary numerical results are confirmed,
this implies that we can, in principle, widen
the throat of a harvested vacuum-defect wormhole
by adding a \emph{finite} amount of \emph{non-exotic} matter.

\begin{figure}[t] 
\vspace*{-0mm}    
\begin{center}
\includegraphics[width=0.75\textwidth]{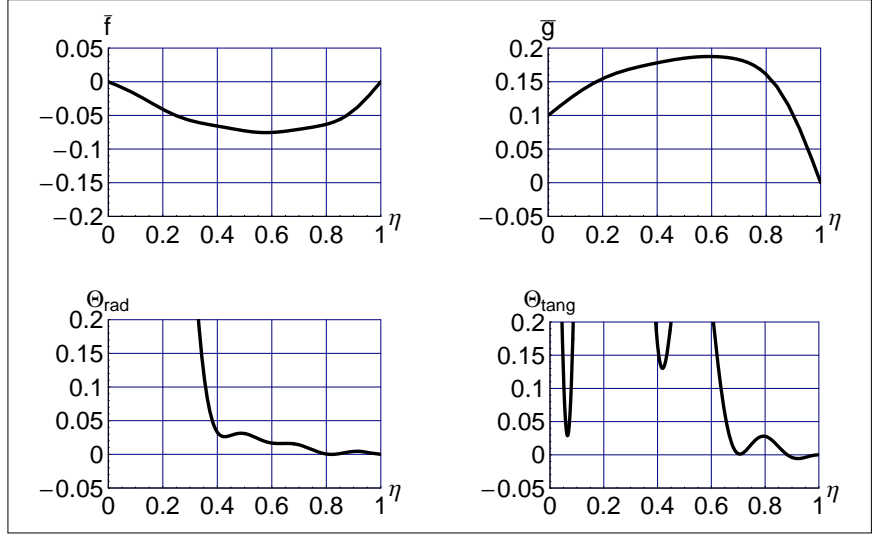}
\end{center}
\vspace*{4mm}
\vspace*{-0mm}
\caption{Numerical results for $N_\text{coeff}=12$
with $\mathcal{P}=6.12347 \times 10^{-6}$ for nonzero coefficients
$c_{1}=-0.0782196$, $c_{2}=0.0107336$, $c_{3}=-0.00647395$, 
$c_{4}=0.00391987$, $c_{5}=-0.00119384$, $c_{6}=0.00229041$
and
$d_{1}=0.151704$, $d_{2}=-0.0321288$, $d_{3}=0.0195948$, 
$d_{4}=-0.00525931$, $d_{5}=0.00290803$, $d_{6}=-0.000818478$.
The $\Theta_\text{rad}$ value   
drops to approximately $-5 \times 10^{-5}$ at $\eta \sim 0.8$
and the $\Theta_\text{tang}$ value to approximately $-5 \times 10^{-3}$ 
at $\eta \sim 0.9$.
}
\label{fig:trend-Ncoeff-12}
\vspace*{500mm}
\end{figure}

\emph{Addendum ---} Using an adapted   
penalty function $\mathcal{P}_\text{NEW}$
and the \texttt{NMinimize} routine of \textsc{Mathematica} 12.1,
we have obtained $N_\text{coeff}=12$ metric functions
\eqref{eq:expansions-fbar-gbar} with a 
penalty value $\mathcal{P} \approx 6.1 \times 10^{-6}$.
The corresponding plots are similar to those
of Fig.~\ref{fig:all-results-Ncoeff-8},
but here we only show Fig~\ref{fig:trend-Ncoeff-12}
as a continuation of the sequence in
Figs.~\ref{fig:trend-Ncoeff-2and4} and \ref{fig:trend-Ncoeff-6and8}.

\end{appendix}

\newpage 

\end{document}